\documentclass[conference]{IEEEtran}
\IEEEoverridecommandlockouts
\usepackage{cite}
\usepackage{amsmath,amssymb,amsfonts}
\usepackage[linesnumbered, ruled,norelsize]{algorithm2e}    
\usepackage{bm}
\usepackage{graphicx}
\usepackage{textcomp}
\usepackage{xcolor}
\ifCLASSOPTIONcompsoc            
  \usepackage[caption=false,font=normalsize,labelfont=sf,textfont=sf]{subfig}
\else
  \usepackage[caption=false,font=footnotesize]{subfig}
\fi
\def\BibTeX{{\rm B\kern-.05em{\sc i\kern-.025em b}\kern-.08em
    T\kern-.1667em\lower.7ex\hbox{E}\kern-.125emX}}

\makeatletter    
\newcommand{\linebreakand}{%
    \end{@IEEEauthorhalign}
    \hfill\mbox{}\par
    \mbox{}\hfill\begin{@IEEEauthorhalign}
}
\newcommand{\removelatexerror}{\let\@latex@error\@gobble}
\makeatother

\begin{document}

\title{Reconfigurable Holographic Surface Aided Wireless Simultaneous Localization and Mapping
}

\author{\IEEEauthorblockN{Haobo~Zhang\IEEEauthorrefmark{1},
Ziang~Yang\IEEEauthorrefmark{1},
Hongliang~Zhang\IEEEauthorrefmark{1},
Boya~Di\IEEEauthorrefmark{1},
and~Lingyang~Song\IEEEauthorrefmark{1}\IEEEauthorrefmark{2}}
\IEEEauthorblockA{\IEEEauthorrefmark{1}School of Electronics, Peking University, Beijing, China.}
\IEEEauthorblockA{\IEEEauthorrefmark{2}School of Electronic and Computer Engineering, Peking University Shenzhen Graduate School, Shenzhen, China.}
}

\maketitle

\begin{abstract}
As a crucial facilitator of future autonomous driving applications, wireless simultaneous localization and mapping~(SLAM) has drawn growing attention recently. However, the accuracy of existing wireless SLAM schemes is limited because the antenna gain is constrained given the cost budget due to the expensive hardware components such as phase arrays. To address this issue, we propose a reconfigurable holographic surface (RHS)-aided SLAM system in this paper. The RHS is a novel type of low-cost antenna that can cut down the hardware cost by replacing phased arrays in conventional SLAM systems. However, compared with a phased array where the phase shifts of parallel-fed signals are adjusted, the RHS exhibits a different radiation model because its amplitude-controlled radiation elements are series-fed by surface waves, implying that traditional schemes cannot be applied directly. To address this challenge, we propose an RHS-aided beam steering method for sensing the surrounding environment and design the corresponding SLAM algorithm. Simulation results show that the proposed scheme can achieve more than there times the localization accuracy that traditional wireless SLAM with the same cost achieves.
\end{abstract}


\section{Introduction}
\label{s_i}

Autonomous driving is a prominent trend in the automotive industry that is widely acknowledged to revolutionize transportation and fuel advancements in sensing, communication, and other technologies. One of the most promising technologies for autonomous driving is wireless simultaneous localization and mapping~(SLAM). By equipping a vehicle with millimeter wave~(mmWave) radars, this technology enables the vehicle to accurately map its surroundings in real-time and obtain its location information for navigation decision-making. Additionally, mmWave radar is more dependable than lidar and camera in inclement weather conditions, making it an ideal option for autonomous driving applications with rigorous safety requirements.

Existing works on wireless SLAM can be classified into two categories based on the utilized antennas, i.e., mechanically~\cite{paul2010radar} or electronically~\cite{petrov2021auto} steerable antennas. In~\cite{paul2010radar}, a mechanically rotating antenna was adopted for the environment perception and the mobile robot localization. Compared with electronically steerable antennae such as a phased array, the mechanically steerable antenna has a simpler structure and higher stability, while the scanning speed of a phased array is faster than that of the mechanically rotating antenna, which is more appropriate to be implemented in fast-moving scenarios. The authors in~\cite{petrov2021auto} introduced a phased array to assist the SLAM system by providing the radar images of the environment, and designed an extended Kalman filter~(EKF) based method for simultaneous localization, mapping, and calibration. 

However, the high hardware costs of phased arrays hinder the further improvement of phased array-assisted SLAM. Specifically, as the SLAM performance is positively related to the sensing accuracy of the radar, there is an increasing pursuit of phased arrays with higher antenna gains and lower estimation errors, which inevitably leads to larger physical apertures and a greater amount of phase shifters. In contrast, existing phased arrays for autonomous driving applications are based on integrated circuit technology, and a single-chip typical has only 4-16 phase shifters~\cite{gu2021packaging}. To fabricate large-scale phase array, a large number of chips need to be connected and synchronized, and the complexity and cost of such a phase array will become unaffordable for commercial applications. 

To address this issue, we propose to use reconfigurable holographic surfaces~(RHSs) rather than large-scale phased arrays to enable SLAM systems. The RHS is a recently developed low-cost metasurface antenna that possesses a vast number of radiation elements. The beam steering of an RHS can be realized by electronically tuning the radiation amplitudes of the series-fed elements, which is able to support a fast scanning rate. Besides, the RHS has a lower cost compared with phased arrays due to a simpler element structure. Specifically, each adjustable element in the RHS only requires one or two active components such as PIN diodes, while a phase shifter typically contains tens of or more transistors~\cite{uxua2023a}, which consumes more power and is more difficult to manufacture.
Due to these advantages, the RHSs has received great attention from both industry and academia. Specifically, Pivotal Commwave has developed commercial RHS prototypes that cover the 5G mmWave frequency band~\cite{pivotal2019holographic}. In the literature, the performance of RHS-aided systems has also been verified for a variety of applications such as communication~\cite{deng2023reconfigurable} and radar detection~\cite{zhang2023holographic}. Due to the benefits of fast scanning rate and low cost, the RHS can also enhance the performance of SLAM systems, while there is no research on the design and evaluation of RHS-aided SLAM schemes.

In this paper, we consider a SLAM scenario where a vehicle performs SLAM tasks with the aid of the equipped RHSs. Specifically, to detect obstacles in different directions, four RHS-aided radars are attached to the four sides of the vehicle, respectively. The received radar signals are processed to retrieve the point clouds for environment mapping and vehicle localization. However, as the surface wave propagation and amplitude-controlled structure in the RHS give rise to a radiation model different from that of a phased array, traditional beam steering technique that adjusts the phase shifters cannot be directly applied, and the point cloud generation method for phased array model has to be redesigned. 

To tackle this challenge, we design an RHS-aided SLAM protocol that regulates the RHS beam scanning during the SLAM process and propose the algorithms for the RHS-aided point cloud generation, localization, and mapping. The effectiveness of the proposed scheme compared with the phased array-based scheme is verified through simulation.

\vspace{-1mm}
\section{System Model}
\label{s_sm}
\vspace{-1mm}
In this section, we first show the considered SLAM scenario, and then describe the models of the RHS-enabled radar.

\vspace{-1mm}
\subsection{Scenario Description}
\label{ss_sd}
\vspace{-1mm}

\begin{figure}[!t]
    \centering
    \includegraphics[height=1.4in]{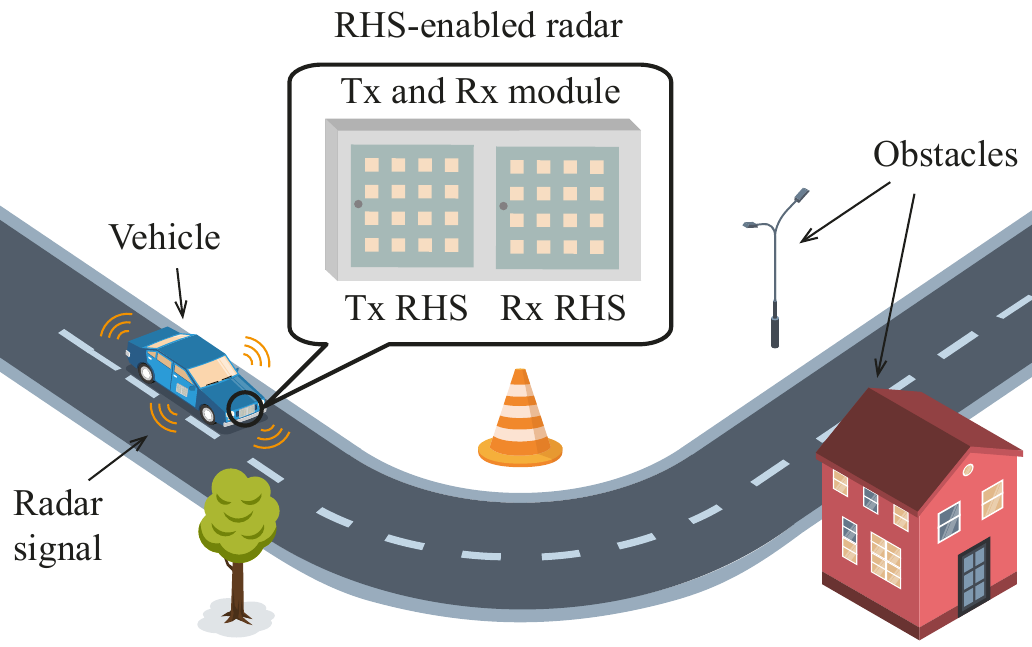}
    \caption{The autonomous driving scenario.}
    \label{f_scenario}
\end{figure}

We consider an autonomous driving scenario consisting of a vehicle and multiple obstacles, as shown in Fig.~\ref{f_scenario}. The vehicle is equipped with four RHS-based radars, each attached on one side of the vehicle to enable 360$^\circ$ azimuth scan. Each radar comprises a transmitter~(Tx), a receiver~(Rx), and two RHSs that serve as Tx and Rx antennas, respectively. The Tx/Rx is connected to the Tx/Rx RHS for signal transmission/reception.

During the SLAM process, the vehicle moves along the road and periodically transmits and receives signals by using its radars. Different beamforming vectors are applied on the RHSs to enable a precise scan of the obstacles. Then, the environment information are extracted from the received signals, which is further analyzed to derive the poses of the vehicle.

\vspace{-1mm}
\subsection{RHS-enabled Radar Model}
\label{ss_rerm}
\vspace{-1mm}

\begin{figure}[!t]
    \centering
    \includegraphics[height=1.4in]{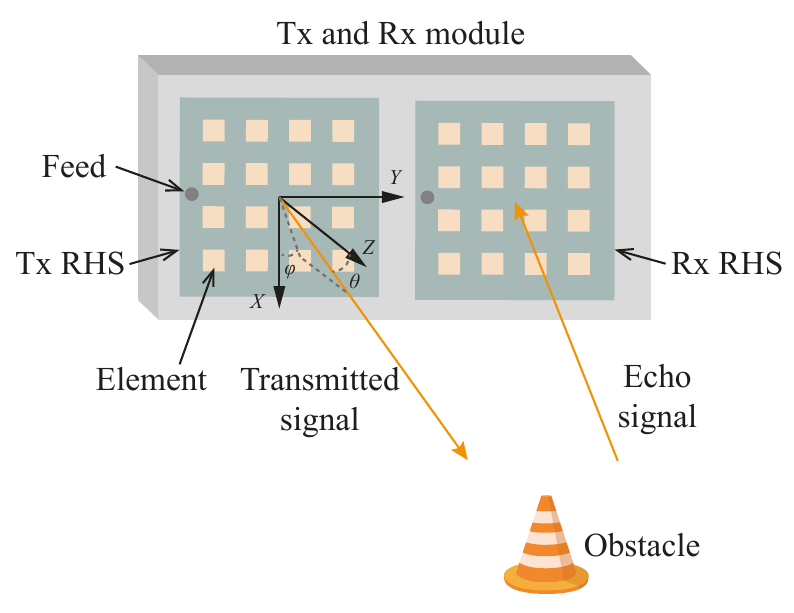}
    \caption{Illustration of the RHS-aided radar.}
    \label{f_radar}
\end{figure}

As shown in Fig.~\ref{f_radar}, the radar detects the obstacles with the help of the Tx and Rx RHSs. Specifically, the RHS is a novel type of metasurface that enables adjustable radiation patterns without the need for complex phase shifters. Typically, an RHS is comprised of a feed and numerous metamaterial elements. The feed is positioned at the bottom or side of the RHS where the radio frequency~(RF) signals are injected. The RF signals propagate along the metamaterial elements as surface waves and excite them in order to emit signals into free space. By varying the bias voltage applied to the element, the amplitude of the signal radiated by the element can be adjusted. In this way the RHS can produce the desired radiation patterns for radar sensing~\cite{zeng2021reconfigurable, zhang2021metalocalization}.

Mathematically, for a Tx RHS, its transmit signal towards a given direction $(\theta, \phi)$ can be given by\vspace{-1mm}
\begin{align}
    y_T(t) = \bm{a}^{\text{T}}_T(\theta, \phi)\bm{\Psi}_T\bm{q}_T x(t),\vspace{-2mm}
\end{align}
where $()^{\text{T}}$ is the transpose operator, $\bm{a}_T(\theta, \phi)$ is the steering vector for Tx RHS towards direction $(\theta, \phi)$, $\bm{\Psi}_T = \text{diag}(\psi_{T, 1}, \cdots, \psi_{T, M_t})$ is the holographic pattern, with $\text{diag}()$ being the diagonal operator and $\psi_{T, m}$ being the radiation amplitude of the $m$-th element. $\bm{q}_T$ is the phase shift vector where the $m$-th entry represents the phase shift from the feed to the $m$-th RHS element. $x(t)$ is the signal at the RHS feed.

Similarly, the expression of the signal received by the feed of the Rx RHS can be given based on the reciprocity of the antenna. Therefore, suppose there are $D$ far-field obstacles, the signal received by the Rx can be given by\vspace{-2mm}
\begin{align}
    y(t) =~& \sum^D_{d=1} \beta_d h(\theta_d, \phi_d) x(t-\tau_d) + \bm{q}^{\text{T}}_R\bm{\Psi}_R\bm{\epsilon}(t),\vspace{-2mm}
\end{align}
where $\beta_d$ is the reflection coefficient of the $d$-th obstacle. $(\theta_d, \phi_d)$ is the direction of the $d$-th obstacle. $h(\theta_d, \phi_d)$ characterizes the overall gain of the Tx and Rx RHSs towards direction $(\theta_d, \phi_d)$, whose expression can be written as\vspace{-2mm}
\begin{align}
    h(\theta_d, \phi_d) =~& \bm{q}^{\text{T}}_R\bm{\Psi}_R\bm{a}_R(\theta_d, \phi_d)\bm{a}^{\text{T}}_T(\theta_d, \phi_d)\bm{\Psi}_T\bm{q}_T,\label{def_h}\vspace{-2mm}
\end{align}
where $\bm{q}_R$ is the phase shift vector of the Rx RHS, $\bm{\Psi}_R$ is the holographic pattern of Rx RHS, $\bm{a}_R(\theta_d, \phi_d)$ is the steering vector of the Rx RHS towards $(\theta_d, \phi_d)$. $\bm{\epsilon}(t)$ is the noise at the Rx RHS elements, and each entry in $\bm{\epsilon}(t)$ follows a complex Gaussian distribution with a mean of $0$ and a variance of $\sigma$.

\vspace{-1mm}
\section{RHS-aided SLAM Protocol}
\label{s_rasp}

\begin{figure}[!t]
    \centering
    \includegraphics[height=1.4in]{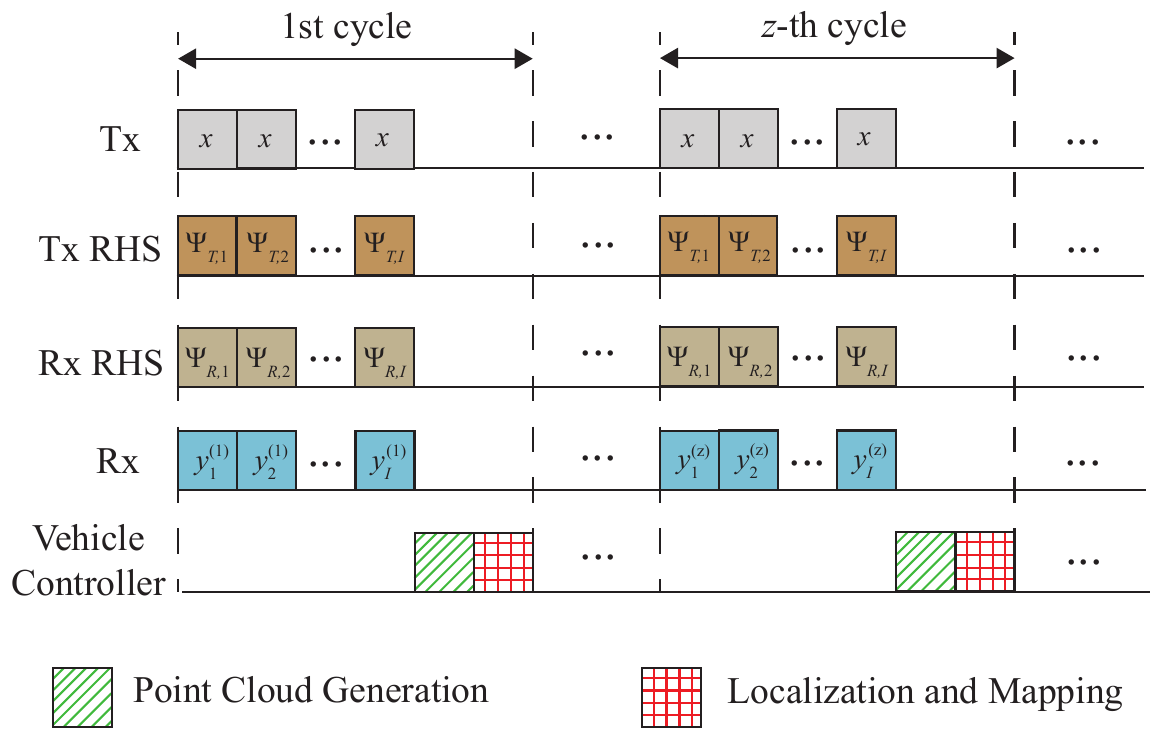}
    \caption{RHS-aided SLAM protocol.}
    \label{f_protocol}
\end{figure}

In this section, an RHS-aided SLAM protocol is designed for the proposed system. The timeline of the whole process is divided into cycles, and three steps are executed sequentially in each cycle, i.e., the radar sensing, point cloud generation, and localization and mapping steps, as illustrated in Fig.~\ref{f_protocol}. The three steps in each cycle is described in the following.

\begin{itemize}
    \item \textbf{Radar Sensing:} At the beginning of each cycle, the vehicle senses the environment with the help of the equipped four radars, which lasts for $\delta_r$ seconds. We divide $\delta_r$ seconds into $I$ time slots. Similarly, the space of interest is also divided into $I$ angular grids, and one grid is scanned in a time slot. Specifically, In the $i$-th time slot of the $z$-th cycle, the Tx transmits signal $x$ to the Tx RHS, whose holographic pattern is set as $\Psi_{T, i}$. At the same time, the holographic pattern of the Rx RHS is adjusted to $\Psi_{R, i}$, and the signal $y^{(z)}_i$ received by the Rx is recorded. The beams of both the Tx and Rx RHSs are steered towards the direction of the $i$-th angular grid, i.e., $(\bar{\theta}_i, \bar{\phi}_i)$. The signal model of this step is introduced in Section~\ref{ss_rerm}, and the optimization of the holographic patterns are provided in Section~\ref{s_rao}.
    \item \textbf{Point Cloud Generation:} In this step, the received signals $\bm{y}^{(z)} = (y^{(z)}_1, \cdots, y^{(z)}_I)$ in the former step are used by the controller of the vehicle to construct the point cloud of the environment, which is denoted by $\mathcal{A}^{(z)}$. The details of this step are given in Section~\ref{s_rpcg}.
    \item \textbf{Localization and Mapping:} The vehicle pose and the environment map are estimated in this step. Let $\hat{\bm{p}}^{(z)}$ and $\hat{\mathcal{M}}^{(z)}$ denote the estimated pose and map, respectively. The pose $\hat{\bm{p}}^{(z)} = (\hat{\bm{t}}^{(z)}, \hat{\bm{r}}^{(z)})^{\text{T}}$ contains the transition $\hat{\bm{t}}^{(z)}$ and rotation $\hat{\bm{r}}^{(z)}$ of the vehicle from the $(z-1)$-th cycle to the $z$-th cycle. Besides, the map $\hat{\mathcal{M}}^{(z)}$ contains the locations of all the obstacles that are estimated based on the signals received in previous $z$ cycles. The localization and mapping algorithm is described in Section~\ref{s_ralm}.
\end{itemize}

\vspace{-2mm}
\section{Holographic Pattern Optimization}
\label{s_rao}
\vspace{-1mm}

In this section, the holographic pattern optimization problem is first formulated, and then the algorithm to tackle the formulated problem is proposed. Note that in order to facilitate the real-time operation of the SLAM system, the optimization of the holographic patterns are performed offline.

\vspace{-1mm}
\subsection{Holographic Pattern Optimization Problem Formulation}
\label{ss_raopfad}
\vspace{-1mm}

In this subsection, the optimization problem for the $i$-th time slot is described. The optimization problem for other time slots can be obtained in a similar way, which are omitted for brevity.

To accurately detect obstacles in $i$-th angular grid and suppress the interference from other directions~\cite{stoica2007on}, we maximize the difference of signal-to-noise ratio~(SNR) between the direction of main lobe ($i$-th angular grid) and those of the side lobes. Thus, the optimization problem can be formulated as
\begin{subequations}
\begin{align}
    \text{P1:}~&\max_{\delta, \bm{\Psi}_{T, i}, \bm{\Psi}_{R, i}} \delta,\label{p1_obj}\\
    s.t.~& ||\bm{\Psi}_{T, i} \bm{q}_T x ||^2_2 = P_M,\label{p1_c1}\\
    & \gamma(\bar{\theta}_i, \bar{\phi}_i) - \gamma(\tilde{\theta}_l, \tilde{\phi}_l) \ge \delta, \forall (\tilde{\theta}_l, \tilde{\phi}_l) \in \mathcal{D},\label{p1_c2}\\
    & \psi_{t, m, i} \in [0, 1], \forall m, i,\label{p1_c3}\\
    & \psi_{r, m, i} \in [0, 1], \forall m, i,\label{p1_c4}
\end{align}
\end{subequations}
where $||\cdot||^2_2$ is the Euclid norm, $P_M$ is the maximum radiation power of the Tx RHS, $\gamma(\bar{\theta}_i, \bar{\phi}_i)$ is the SNR of the signal reflected from direction $(\bar{\theta}_i, \bar{\phi}_i)$. Based on~(\ref{def_h}), we have

\begin{align}
    \gamma(\bar{\theta}_i, \bar{\phi}_i) = \dfrac{|h(\bar{\theta}_i, \bar{\phi}_i)|^2}{\bm{q}^{\text{T}}_R\bm{\Psi}_{R, i}\bm{\Psi}_{R, i}\bm{q}^*_R\sigma^2}.\vspace{-1mm}
\end{align}
$(\tilde{\theta}_l, \tilde{\phi}_l)$ is a side lobe direction. $\mathcal{D}$ is a discrete set that covers the entire angular region of side lobes. $\psi_{t, m, i}$ and $\psi_{r, m, i}$ are the $(m, m)$-th entry of $\bm{\Psi}_{T, i}$ and $\bm{\Psi}_{R, i}$, respectively. Constraint~(\ref{p1_c1}) sets the radiation power of Tx RHS. Constraint~(\ref{p1_c2}) restricts the side lobe levels of the radar, and constraints~(\ref{p1_c3}) and (\ref{p1_c4}) limit the values of the radiation amplitudes.

Since the variables $\delta$, $\bm{\Psi}_{T, i}$, and $\bm{\Psi}_{R, i}$ are coupled in~(P1), which is non-trivial, we decouple (P1) into two subproblems, i.e., transmit and receive pattern optimization subproblems.

\subsubsection{Transmit Pattern Optimization Subproblem}

Given the holographic pattern $\bm{\Psi}_{R, i}$ of Rx RHS, the transmit subproblem can be formulated as\vspace{-1mm}
\begin{align}
    \text{P2:}~&\max_{\delta, \bm{\Psi}_{T, i}} \delta, s.t. \text{(\ref{p1_c1})}, \text{(\ref{p1_c2})}, \text{(\ref{p1_c3})}.\vspace{-1mm}
\end{align}

\subsubsection{Receive Pattern Optimization Subproblem}

Given the holographic pattern $\bm{\Psi}_{T, i}$ of Tx RHS, the receive subproblem can be formulated as\vspace{-1mm}
\begin{align}
    \text{P3:}~&\max_{\delta, \bm{\Psi}_{R, i}} \delta, s.t. \text{(\ref{p1_c2})}, \text{(\ref{p1_c4})}.\vspace{-1mm}
\end{align}

\subsection{Holographic Pattern Optimization Algorithm}
\label{ss_raoa}

In this subsection, we design the holographic pattern optimization algorithm that iteratively solves the above two subproblems to optimize the variables $\delta$, $\bm{\Psi}_{T, i}$, and $\bm{\Psi}_{R, i}$. To be specific, the holographic pattern $\bm{\Psi}_{R, i}$ is first randomly initialized. Next, in each iteration, subproblems (P2) and (P3) are sequentially solved. The iteration ends when the values of $\delta$ in two adjacent iterations is less than $\delta_t$, a pre-determined threshold. The algorithms to solve subproblems (P2) and (P3) in each iterations are provided as follows.

\subsubsection{Transmit Pattern Optimization Algorithm}

Since the objective function and constraint (\ref{p1_c3}) in (P2) are linear, and constraints (\ref{p1_c1}) and (\ref{p1_c2}) are quadratic functions of $\bm{\Psi}_{T, i}$, (P2) is a quadratic constrained linear program and can be efficiently solved by existing methods such as semi-definite relaxation.

\subsubsection{Receive Pattern Optimization Algorithm}

In this part, we show how to obtain the optimal solution of (P3). Specifically, let $\delta^{(1)}$ and $\bm{\Psi}^{(1)}_{R, i}$ represent the optimal solution of (P3). Besides, define $\bm{\Psi}^{(2)}_{R, i} = x^{(1)} \bm{\Psi}^{(1)}_{R, i}$ that satisfies $||\bm{\Psi}^{(2)}_{R, i} \bm{q}_T||^2_2\sigma^2 = 1$. We can find that $\delta^{(1)}$ and $\bm{\Psi}^{(2)}_{R, i}$ is a feasible solution of the following problem:\vspace{-1mm}
\begin{subequations}
    \begin{align}
        \text{P4:}~&\max_{\delta, \bm{\Psi}_{R, i}} \delta,\label{p4_obj}\\
        s.t.~& ||\bm{\Psi}_{R, i} \bm{q}_T||^2_2\sigma^2 = 1,\label{p4_c1}\\
        & |h(\bar{\theta}_i, \bar{\phi}_i)|^2 - |h(\tilde{\theta}_i, \tilde{\phi}_i)|^2 \ge \delta, \forall (\tilde{\theta}_l, \tilde{\phi}_l) \in \mathcal{D}.\label{p4_c2}\vspace{-1mm}
    \end{align}   
\end{subequations}
This is because $\delta^{(1)}$ and $\bm{\Psi}^{(1)}_{R, i}$ satisfies (\ref{p1_c2}), which guarantees that (\ref{p4_c2}) is satisfied. Actually, (P4) is a linear program with quadratic constraints, which can also be efficiently solved. Let $\delta^{(3)}$ and $\bm{\Psi}^{(3)}_{R, i}$ denote the optimal solution of (P4), and we have $\delta^{(3)} \ge \delta^{(1)}$. Similarly, we can show that $\delta^{(3)}$ and $\bm{\Psi}^{(4)}_{R, i}$ is a feasible solution of (P3), where $\bm{\Psi}^{(4)}_{R, i} = x^{(3)} \bm{\Psi}^{(3)}_{R, i}$ that satisfies~(\ref{p1_c4}). Thus, we have $\delta^{(3)} \le \delta^{(1)}$. As a result, $\delta^{(3)} = \delta^{(1)}$, and the optimal solution of (P3) is derived.

\section{RHS-aided Point Cloud Generation}
\label{s_rpcg}

In this section, we propose the RHS-aided point cloud generation algorithm that uses the optimized holographic patterns to derive the point cloud of the environment. The algorithm consists of two phases, i.e., range estimation and angle estimation phases. 
In the first phase, the ranges of the obstacles in a given angular grid are first estimated based on the signal received in each time slot. Next, in the second phase, for each detected obstacle, its direction is estimated based on the compressed sensing method.

\subsection{Range Estimation}
\label{ss_re}

To detect obstacles towards direction $(\bar{\theta}_i, \bar{\phi}_i)$ and estimate their ranges, the convolution between $y^{(z)}_i(t)$ and matched filter $x^*(t)$ is first calculated, where $()^*$ is the conjugate operator. The expression of the convolution is given by
\begin{align}
    \iota_i(\tau) = \int y^{(z)}_i(t)x^*(t - \tau) dt.
\end{align}
The local maxima point $\tau_n$ that satisfies $\iota_i(\tau_n) > \delta_d$ is recognized as the delay of an obstacle, whose range is $\tau_n c/2$. Let $\mathcal{O}$ denote the set that contains all the local maxima points related to the signals received in all the time slots.

\subsection{Angle Estimation}
\label{ss_ae}

In this part, the angles of the detected obstacles are estimated by using the compressed sensing method to improve the estimation accuracy. Specifically, for a delay $\tau_n$ in $\mathcal{O}$, we use the measurement vector $\bm{\iota}_n = (\iota_1(\tau_n), \cdots, \iota_I(\tau_n))^{\text{T}}$ to estimate the number and the directions of the targets. According to the principle of compressed sensing, the estimation problem can be formulated by
\begin{subequations}
\begin{align}
    \text{P5:}~&\min_{\bm{s}} ||\bm{s}||_0,\\
    s.t.~& \bm{\iota}_n = \bm{H} \bm{s},
\end{align}
\end{subequations}
where $||\cdot||_0$ is the $l_0$ norm, $\bm{H}$ is a $I \times I_c$ matrix, and the $(i, i')$-th entry in $\bm{H}$ is
\begin{equation}
    h_{i, i'} = \bm{q}^{\text{T}}_R\bm{\Psi}_{R, i}\bm{a}_R(\theta_{i'}, \phi_{i'})\bm{a}^{\text{T}}_T(\theta_{i'}, \phi_{i'})\bm{\Psi}_{T, i}\bm{q}_T.
\end{equation}
$\bm{s}$ is the obstacle vector to be constructed. If the $i'$-th entry of $\bm{s}$, denoted by $s_i$, is not zero, it means that these exists an obstacle in direction~$(\theta_{i'}, \phi_{i'})$.

To tackle (P5) that is NP-hard, we alter the orthogonal matching pursuit~(OMP) method to accommodate to the radiation model of the RHS. To initialize, the residual vector $\bm{\iota}_{res}$ is set as $\bm{\iota}_n$, and the obstacle vector is set as $\bm{0}$. Next, in each iteration, the algorithm first find a direction $i'$ that has the best match with the residual, i.e., $|\bm{h}^{\text{H}}_{i'} \bm{\iota}_{res}| \ge |\bm{h}^{\text{H}}_{i''} \bm{\iota}_{res}|, \forall i'' \ne i'$. Then, based on the least squares method, the estimate of the obstacle in this direction can be given by $s_{i'} = \bm{h}^{\text{H}}_{i'} \bm{\iota}_{res} / ||\bm{h}_{i'}||^2_2$. The measurement related to this direction, i.e., $\bm{h}_{i'}s_{i'}$ is deleted from the residual $\bm{\iota}_{res}$. The iteration terminates when the number of iterations, i.e., $l$, is larger than $I$, or the power of the residual is less than $\delta_r$. Thus, the point cloud $\mathcal{A}^{(z)}$ that contains all the locations of the obstacles are derived.

\begin{figure}[!t]
    \removelatexerror
    \begin{algorithm}[H]
      \caption{RHS-aided Point Cloud Generation Algorithm}
      \label{a_aea}
      \KwIn{Received signal $\bm{y}^{(z)}$;}
      \KwOut{Point cloud $\bm{A}^{(z)}$;}
      Initial $\mathcal{O} = \varnothing$\;
      \For{each angular grid $(\bar{\theta}_i, \bar{\phi}_i)$}{
        Calculate the convolution $\iota_i(\tau)$ between $y^{(z)}_i(t)$ and the matched filter $x^*(t)$\;
        Add the local maxima point $\tau_n$ of $\iota_i(\tau)$ into $\mathcal{O}$\;
      }
      \For{$\tau_n \in \mathcal{O}$}{
        Compute measurement vector $\bm{\iota}_n$ and matrix $\bm{H}$\;
        Set residual $\bm{\iota}_{res} = \bm{\iota}_n$, vector $\bm{s} = \bm{0}$, and $l = 1$\;
        \While{$l \le I$ or $||\bm{\iota}_{res}||^2_2 \ge \delta_r$}{
            Find a column of $\bm{H}$, denoted by $\bm{h}_{i'}$ that maximizes $|\bm{h}^{\text{H}}_{i'} \bm{\iota}_{res}|$\;
            Modify $s_{i'} = \bm{h}^{\text{H}}_{i'} \bm{\iota}_{res} / ||\bm{h}_{i'}||^2_2$\;
            Update residual $\bm{\iota}_{res} = \bm{\iota}_{res} -\bm{h}_{i'}s_{i'}$\;
            Set $l = l + 1$\;
        }
      }
    \end{algorithm}
\end{figure}

\section{RHS-aided Localization and Mapping}
\label{s_ralm}

In this section, we propose the localization and mapping algorithm that converts the RHS-aided point cloud to the estimation of the vehicle pose and the environment. There are three phases in the algorithm, including the RHS-aided point cloud registration, data association, and information update phases. The transition and rotation of the vehicle between two adjacent cycles are first calculated in the first phase. The obstacles in the point cloud can then be matched with the obstacles in the environment in the second phase. Finally, the vehicle pose and the map are updated in the last phase.

\subsection{RHS-aided Point Cloud Registration}
\label{ss_pcr}

A convolutional neural network~(CNN)-based algorithm is designed for accurate point cloud registration. Specifically, the point clouds $\mathcal{A}^{(z)}$ and $\mathcal{A}^{(z-1)}$ are first transformed as grid maps~\cite{Mafukidze2022scattering} as the input of the CNN, and the algorithm outputs the estimated transition $\hat{\bm{t}}^{(z)}$ and rotation $\hat{\bm{r}}^{(z)}$ of the vehicle between the $(z-1)$-th and the $z$-th cycle. The CNN has four layers, i.e., a convolutional layer, a max-pooling layer, and two fully connected layers. The CNN is trained by minimizing the following loss function:
\begin{align}
    L(\bm{w}, \bm{v}) =~& \dfrac{1}{N_b}\sum^{N_b}_{n=1}\biggl(||\hat{\bm{t}}^{(z)}_n - \bm{t}^{(z)}_n||^2_2 + \gamma_r||\hat{\bm{r}}^{(z)}_n - \bm{r}^{(z)}_n||^2_2\biggr),
\end{align}
where $\bm{w}$ is the parameters in the CNN model, $\bm{v}$ is the training data, $N_b$ represents the batch size, $\hat{\bm{t}}^{(z)}_n$ and $\hat{\bm{r}}^{(z)}_n$ are the estimated transition and rotation based on the $n$-th batch of the training data, respectively. $\bm{t}^{(z)}_n$ and $\bm{r}^{(z)}_n$ are the corresponding ground truth. $\gamma_r$ is a pre-determined weighting factor.

\vspace{-1mm}
\subsection{Data Association}
\label{ss_da}

The obstacles in the point cloud $\mathcal{A}^{(z)}$ are matched to those in the map $\hat{\mathcal{M}}^{(z-1)}$ based on the predicted transition $\hat{\bm{t}}^{(z)}$ and rotation $\hat{\bm{r}}^{(z)}$. Specifically, the data association contains the following two steps:
\begin{itemize}
    \item \textbf{Position Compensation:} As the map and the point cloud are in different coordinates, the locations of the obstacles in the point cloud, where the vehicle is at the origin of the coordinate, needs to be transformed to those in the map, where the starting point of the vehicle trajectory is at the origin.
    \item \textbf{Cluster Match:} In this step, each obstacle in the point cloud will be matched to a obstacle with the lowest relative distance in the map. However, if this relative distance is larger than a threshold $\delta_m$, this obstacle in the point cloud will not be matched to any obstacle in the map.
\end{itemize}

\subsection{Information Update}
\label{ss_iu}

\begin{figure}[!t]
    \centering
    \includegraphics[width=3in]{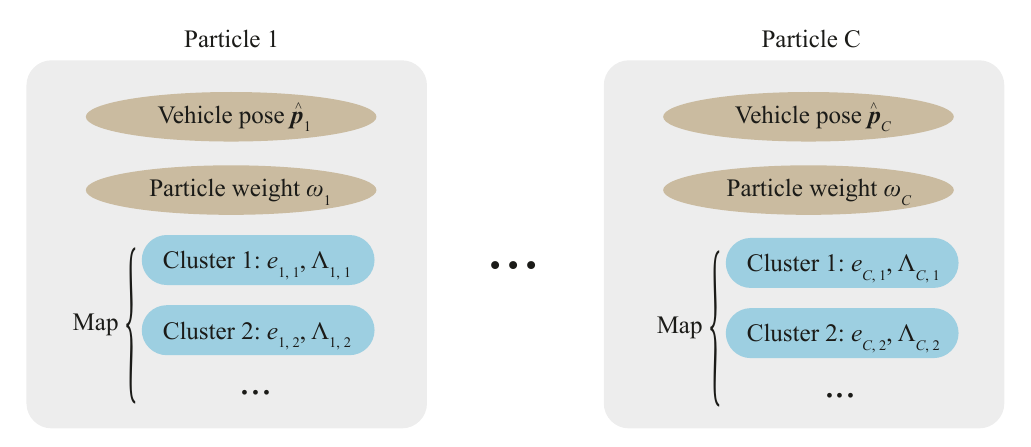}
    \caption{Particle filter in the information update phase.}
    \label{f_particle}
\end{figure}

In this phase, a Rao-Blackwellized particle filter~(RBPF)-based algorithm~\cite{yang2022metaslam, hua2023intelligent} is designed to update the vehicle pose $\bm{p}^{(z)}$ and map $\hat{\mathcal{M}}^{(z)}$ in the $z$-th cycle. The algorithm represents the possible vehicle poses by using a set of particles with different weights, as shown in Fig.~\ref{f_particle}. Specifically, in each particle, there is a vehicle pose, particle weight, and the estimation of the obstacle positions. The location of the $u$-th obstacle in the $c$-th particle is characterized by a Gaussian distribution with mean $\bm{e}_{c, u}$ and variance $\bm{\Lambda}_{c, u}$. The estimation of the vehicle pose can be calculated by weighted summing all the poses in the particles, and the estimation of the obstacle location can be derived in a similar way.

\begin{figure*}[!t]
    \centering
    \subfloat[]{
        \includegraphics[height=1.75in]{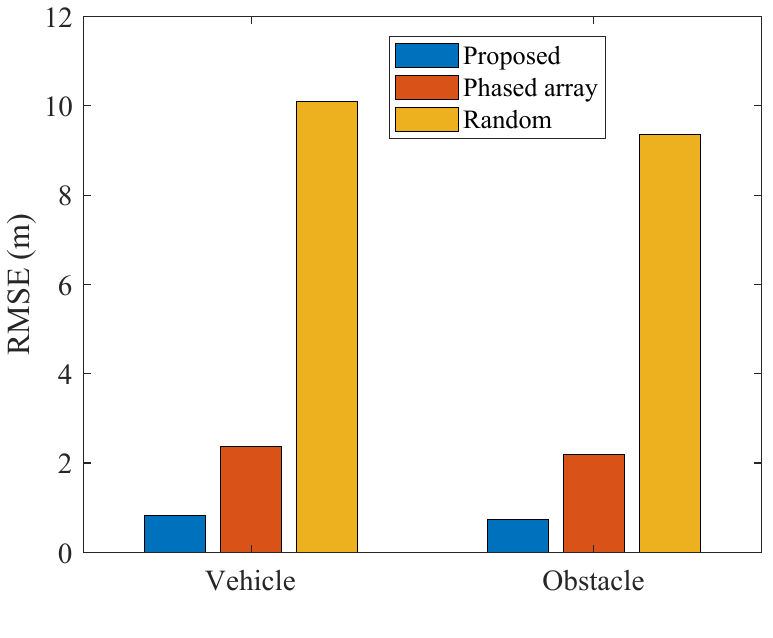}
    }
    \hspace{0.1in}
    \subfloat[]{
      \includegraphics[height=1.75in]{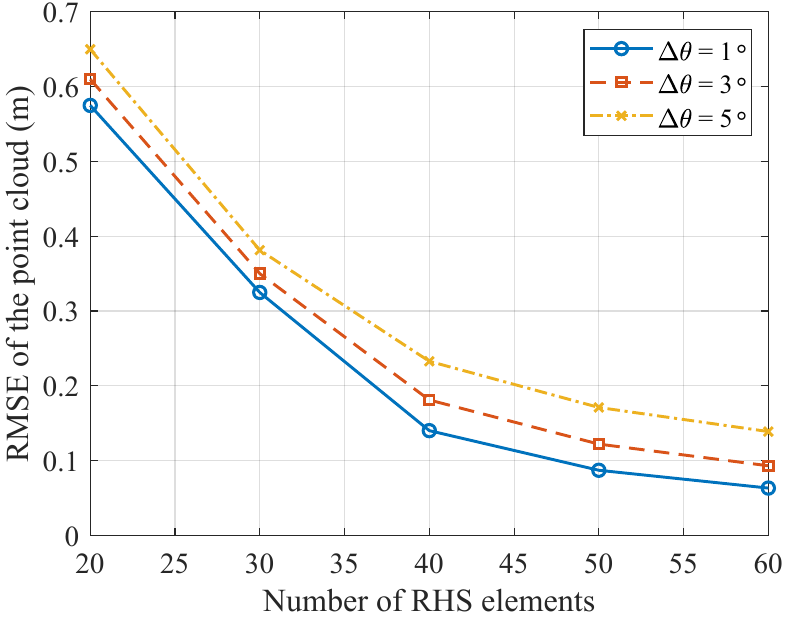}
    }
    \hspace{0.1in}
    \subfloat[]{
      \includegraphics[height=1.75in]{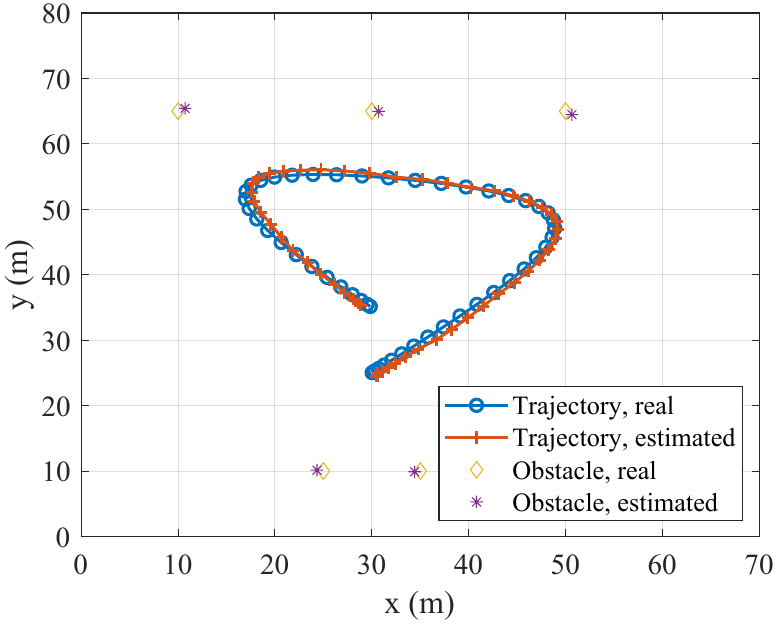}
    }    
    \caption{(a) The RMSE of the estimated locations of the vehicle and the obstacles; (b) The RMSE of the estimated locations in the point cloud versus the number of RHS elements; (c) Illustration of the estimated trajectory and the map.}
    \label{f_simulation}
    \vspace{-3mm}
\end{figure*}

The point cloud $\mathcal{A}^{(z)}$ and the data association relationship derived in the above phases can be used to update the weights and locations in the particles, thus updating the vehicle pose and the map. Specifically, the information updating has the following four steps:
\begin{itemize}
    \item \textbf{Pose and Location Update:} The vehicle pose in each particle is first updated by using the white noise acceleration model. Next, the location of the $u$-th obstacle in the map of the $c$-th particle is updated by
    \begin{align}
        \bm{e}_{c, u} \leftarrow~& \bm{e}_{c, u} + \bm{\Lambda}_{c, u}\bm{F}^{\text{T}}_{u'} \bm{J}^{-1}_{u'}(\hat{\bm{p}}_c - \bm{p}_{u'}),\\
        \bm{\Lambda}_{c, u} \leftarrow~& (\bm{I} - \bm{\Lambda}_{c, u}\bm{F}^{\text{T}}_{u'} \bm{J}^{-1}_{u'}\bm{F}_{u'})\bm{\Lambda}_{c, u},
    \end{align}

    where $u'$ is the index of the obstacle in the point cloud that associates with the $i$-th obstacle in the particle, $\bm{F}_{u'}$ is the jacobian matrix of function $\bm{f}_{u'}$ that predicts the pose $\bm{p}_{u'}$ given $\bm{e}_{c, u}$ and the vehicle pose $\hat{\bm{p}}_c$. $\bm{I}$ is the identity matrix, and $\bm{J}$ can be expressed as
    \begin{equation}
        \bm{J}_{u'} = \bm{F}_{u'} \bm{\Lambda}_{c, u} \bm{F}^{\text{T}}_{u'} + \bm{J}_c,
    \end{equation}
    where $\bm{J}_c$ is the covariance matrix of the estimation noise.
    \item \textbf{Weight Calculation:} In this step, the weight of each obstacle in a particle is adjusted as the reflection amplitude of this obstacle. This is because a higher reflection amplitude leads to a higher location estimation accuracy of the obstacle. Then, the weight of each particle is derived by multiplying the weights of the obstacles in the particle together.
    \item \textbf{Resampling:} The particles are resampled in this step, and the resampling rate of each particle is proportional to the weight of this particle in order to keep the particles with poses close that of the vehicle and reduce the number of other particles.
    \item \textbf{Vehicle Pose and Map Update:} The vehicle pose $\hat{\bm{p}}^{(z)}$ can be calculated by weighted summing all the poses in the particles. Similarly, the obstacle locations in the map $\hat{\mathcal{M}}^{(z)}$ can also be derived.
\end{itemize}

\section{Simulation Results}
\label{s_sr}

In this section, the simulation results of the proposed RHS-aided SLAM system is provided. As discussed in Section~\ref{ss_sd}, the vehicle is equipped with four radars, and each radar has two RHSs for signal transmission and reception. The frequency modulated continuous wave~(FMCW) is applied as the radar waveform. The frequency band of the RHS-based radars in the vehicle is $24 - 24.25$GHz. Each RHS in the vehicle has $60$ elements, and the element spacing is a quarter of the carrier wavelength. The radiation power of the RHS-based radar is $30$dBm, and the noise is set as -$40$dBm.

Fig.~\ref{f_simulation}(a) depicts the comparison between the proposed scheme with two other schemes, i.e., the phased array-based scheme and the random scheme. Specifically, in the former scheme, the RHSs are replaced by the phased array with the same cost\footnote{Based on~\cite{pivotal2019holographic}, we assume that the cost of a phased array element is six times as much as that of an RHS element.}. And in the latter scheme, the patterns of the RHSs are randomly selected. We can find that the RMSEs of both the vehicle and the obstacles obtained by the proposed scheme are about $1.5$m lower than that obtained by the phased array-based scheme, which shows the superiority of the proposed scheme. Besides, the RMSEs obtained by the proposed scheme are much smaller than that obtained by the random scheme. This is because by optimizing the holographic patterns, the echo signal in the target direction is amplified, and the interference signals from other directions are suppressed, which improves the accuracy of the point cloud.

Fig.~\ref{f_simulation}(b) shows the accuracy of the point cloud obtained by the proposed RHS-enabled radar. The root mean square error~(RMSE) of the estimation locations in the point cloud are selected as the performance metric. We can observe that the RMSE decreases when the number of RHS elements increases. This is because a higher antenna gain can be obtained by using an RHS with larger size. Besides, when decreasing the size of the angular grid $\Delta \theta$ from $5^\circ$ to $1^\circ$, the RMSE also decreases. However, the number of angular grids that are scanned in the radar sensing steps in each SLAM cycle will increase, which indicates a tradeoff between the time cost of the radar sensing and the estimation accuracy.

Fig.~\ref{f_simulation}(c) illustrates the simulation results of the vehicle trajectory and the map. Five obstacles located at $(10, 65)$m, $(25, 10)$m, $(30, 65)$m, $(35, 10)$m, and $(50, 65)$m are set in the environment. Besides, there are $50$ cycles in the SLAM process. We can observe that the estimated trajectory and the locations of the obstacles are close to the real ones, which verifies the effectiveness of the proposed SLAM scheme.

\section{Conclusion}
\label{s_c}

In this paper, we have considered the RHS-aided SLAM scenario, where the moving vehicle locates itself and senses the environment by using the equipped RHSs. An RHS-aided SLAM protocol has been designed to regulate the RHS beam scanning in the SLAM process. To generate high-precision point clouds, we have design the algorithm to optimize the holographic patterns, and proposed point cloud generation algorithm to derive the point cloud from the signals received by the RHSs. A localization and mapping algorithm has also been designed to estimate the vehicle pose and the map from the point cloud. Simulation results have shown that: 1) the estimation error of the vehicle trajectory obtained by the proposed scheme is one-third of that obtained by the phased array-based scheme with the same hardware cost; 2) the point cloud estimation error in the proposed scheme can be reduced by increasing the number of RHS elements or scanning the space of interest with a higher angular resolution.

\bibliographystyle{IEEEtran}
\bibliography{IEEEabrv,myReference}

\end{document}